\documentclass[prb,preprint,letterpaper,noeprint,longbibliography,nodoi,footinbib]{revtex4-1} 

\usepackage[colorlinks, allcolors=blue]{hyperref}
\usepackage{amsmath,amssymb}  
\usepackage{amsfonts} 
\usepackage{graphicx} 

\begin{document}


\title{A unified treatment of the redshift, the Doppler effect, and the time dilation 
in general relativity}

%
\author{Masumi Kasai}
\email{kasai@hirosaki-u.ac.jp} 
\affiliation{Graduate School of Science and Technology, Hirosaki University, Hirosaki 036-8561, Japan}


\date{\today}

\begin{abstract}
    We present a unified treatment of the gravitational and cosmological redshift, the Doppler effect due to the moving observer or light source, and the time dilation in the gravitational field in the framework of general relativity.  
    The primary purpose of this paper is to extend the description of Narlikar (1994) on the unified approach towards the redshifts and the Doppler effect in a more generalized form, with the help of the four facts extracted from the comprehensive review article by Ellis (1971).   
    We apply it to the cases of moving observer or light source in the gravitational field and obtain the Doppler effect term, in addition to the standard gravitational or cosmological redshift.  
    The secondary purpose is to explicitly show that the time dilation of a moving clock in the gravitational field can also be understood within the same framework of the unified treatment.  
    We examine the time dilation of the moving clock on geodesic in the gravitational field.  
    We also derive the time dilation of the moving clock on elliptical orbit, based on the same unified treatment.  
    The tertiary purpose is to show that we can understand special-relativistic effects without using the Lorentz transformation.  
    We derive the special-relativistic formulae such as the Doppler effect and aberration of light, the kinetic time dilation, and the Lorentz contraction in the general-relativistic framework.  
\end{abstract}


\maketitle

\section{Introduction}

The Doppler effect of light is usually explained using the Lorentz transformation in special relativity.  
In the presence of gravity, however, the Lorentz transformation does not hold.  

Narlikar\cite{nar} described a unified approach to bring the gravitational and cosmological redshifts within the same framework as the Doppler shift.  (See also 
Schr\"{o}dinger\cite{sch} and Synge\cite{Synge}.)  

In this paper, we extend the approach by Narlikar\cite{nar} in a more generalized form, with the help of the four facts extracted from the comprehensive review article by Ellis\cite{ellis}, which was not referred by Narlikar\cite{nar}.  The description of Ellis\cite{ellis} is originally intended to apply to observations in cosmological models.  Thanks to the covariant feature of the description, however, it can also be applied to other situations, including the gravitational redshift in the Schwarzschild spacetime and even to the purely special relativistic situations.  

The unified treatment is simply based on the following facts (which are extracted from the comprehensive review article by Ellis\cite{ellis}): 
\begin{itemize}
    \item[I.]  Light obeys the null geodesic equation, i.e., the propagation 4-vector 
    $k^{\mu}$ satisfies $$k^{\mu}_{\ ;\nu} k^{\nu} = 0\,, \quad k_{\mu} k^{\mu} = 0\, ,$$
    which will be summarized in Sec.~\ref{sec:ngeflr}.
    \item[II.]  The frequency of light measured by an observer with 4-velocity $u^{\mu}$ is 
    $$\omega = - k_{\mu} u^{\mu}\ ,$$
    which will be explained in Sec.~\ref{sec:decolight}.
    \item[III.]  The 4-velocities of two observers with relative velocity $V$ are related  by     
    $$\bar{u}^{\mu} = \frac{u^{\mu} + V e^{\mu}}{\sqrt{1-V^2}}, \quad e_{\mu} e^{\mu} = 1, 
    \quad e_{\mu} u^{\mu} = 0 , $$
    which will be explained in Sec.~\ref{sec:comporule}.  
    \item[IV.]  The propagation 4-vector of light $k^{\mu}$ can be decomposed into 
    $$k^{\mu} = \omega \left(u^{\mu} + \gamma^{\mu}\right), \quad 
    \gamma_{\mu} \gamma^{\mu} = 1, \quad  \gamma_{\mu} u^{\mu} = 0 , $$
    which will be explained in Sec.~\ref{sec:decolight}.  
\end{itemize}

In the descriptions by Narlikar\cite{nar} and Ellis\cite{ellis}, facts I and II are assumed.   
Facts III and IV, which are described in the article by Ellis\cite{ellis}, are the keys to extend the unified treatment.  

In order to understand the Doppler effect, the gravitational and the cosmological redshifts in a unified manner, the four facts I-IV are the essential and definitive tools.  
Since the Doppler effect and the redshifts are used so often in different contexts and in different interpretations (see Refs.~5 and 6 for example), it is worth again appreciating the unified treatment of them according to the four facts described by Ellis\cite{ellis}.  This is the primary purpose of this paper.  

After extracting the four facts I-IV from the comprehensive review article by Ellis\cite{ellis} in a slightly simplified form, 
we utilize them to derive the basic formulae for the Doppler effect and the redshift in a unified manner.  

The main advantages of our approach over Narlikar's\cite{nar} are as follows.  
First, with the help of the facts I-IV, we can clarify the definitions, the equivalence, and the distinctions  of the frequency shifts in the following way:  
The redshift $z$ of a source as measured by a ``standard'' observer with 4-velocity $u^{\mu}$ is defined by the ratio of the two frequencies at the different points in the spacetime: 
$$1 + z  = \frac{(k_{\mu} u^{\mu})_{\rm source}}{(k_{\mu} u^{\mu})_{\rm observer}}\ . $$
If the light, denoted by the 4-vector $k_{\mu}$, is propagating through the static spacetime, say, the Schwarzschild, $z$ is called the gravitational redshift.  
If the light is propagating through the expanding universe, the same $z$ is called the cosmological redshift.  
On the other hand, the Doppler effect is expressed by the following ratio of the two frequencies at the same point in the spacetime: 
\begin{equation*}
    \frac{(k_{\mu} \bar{u}^{\mu})_{\rm source}}{(k_{\mu} u^{\mu})_{\rm source}}
        \quad\mbox{or}\quad
    \frac{(k_{\mu} u^{\mu})_{\rm observer}}{(k_{\mu} \bar{u}^{\mu})_{\rm observer}} \, .
\end{equation*}
Those will be explained in Sec.~\ref{sec:red} and \ref{sec:Dop}.

Second, although Narlikar\cite{nar} only considered the radial motion, thus only the longitudinal Doppler effects, combining facts III and IV enables us to show that the well-known formulae for the Doppler effect and the aberration of light 
$$ \bar{\omega} = \omega \frac{\sqrt{1-V^2}}{1 - V \cos\bar{\vartheta}}, \quad
\cos\bar{\vartheta} = \frac{\cos\vartheta + V}{1 + V \cos\vartheta}$$
hold for any incident angle $\vartheta$ in both special and general relativity, which will be explained in Sec.~\ref{sec:Dop}.  
It should be noted that the Lorentz transformation is not required to derive these formulae.

The secondary purpose of this paper is to explicitly show that the time dilation of a moving clock in the gravitational field can also be understood within the same framework of the unified treatment.  
Since the time dilation is closely related to the gravitational redshift, it is instructive to show that all we need is the four facts I-IV to understand the time dilation in special and general relativity.  

The tertiary purpose is to show that we can understand special-relativistic effects without using the Lorentz transformation.  
Thanks to the covariant formulation of the basic principles and equations presented in this paper, they hold also in special-relativistic situations.  
Therefore, we can derive special-relativistic formulae such as the Doppler effect and aberration of light, the kinetic time dilation (often called as the special-relativistic time dilation), and the Lorentz contraction in the general-relativistic framework. 

We apply the unified treatment to the cases of moving observer or light source in the gravitational field, and investigate the Doppler effect due to the moving observer or source.  

We also investigate the time dilation of a moving clock on geodesic in the gravitational field. 
We examine the time dilation of the the moving clocks on radial orbit, circular orbit, and non-circular elliptical orbit. 
We use the unit $c = 1$. Greek indices run from $0$ to $3$.  

\section{Basic principles and equations}

We briefly summarize the basic principles and equations for light ray observation in the gravitational field.   
Most of them are described in the review article by Ellis\cite{ellis}. 

\subsection{The null geodesic equation for light rays}\label{sec:ngeflr}

Let us define the propagation 4-vector $k^{\mu}$.  
The light rays whose tangent vector is $k^{\mu}$ are null geodesics \cite{ellis69}: 
\begin{eqnarray}
    k^{\mu} &\equiv& \frac{dx^{\mu}}{dv}, \\
    k^{\mu}_{\ ;\nu} k^{\nu} &=& 0 \,, \label{eq:geode}\\
    k_{\mu} k^{\mu} &=& 0\, ,
\end{eqnarray}
where $v$ is an affine parameter along the null geodesic.  
It is sometimes more convenient to use the geodesic equation for the covariant components 
$k_{\mu} = g_{\mu\nu} k^{\nu}$.  
From $k_{\mu;\nu} k^{\nu} =0$, we obtain
\begin{equation}
    \frac{d k_{\mu}}{dv} = \frac{1}{2} g_{\alpha\beta, \mu} k^{\alpha} k^{\beta}\,. \label{eq:geodek}
\end{equation}
It is convenient in the following sense: if the metric does not depend on some coordinate, say, $x^0$, Eq.~\eqref{eq:geodek} immediately shows us the existence of the conserved quantity:
\begin{eqnarray}
    \mbox{if}\ \ g_{\alpha\beta, 0} &=& 0\,, \\
    \mbox{then}\ \ \frac{d k_0}{dv} &=& \frac{1}{2} g_{\alpha\beta, 0} k^{\alpha} k^{\beta} = 0\,, \\
    \therefore\ \ k_0 &=& \mbox{const.}
\end{eqnarray}

\subsection{The geodesic equation for observers}

Let us define the tangent 4-vector $u^{\mu}$ of the world line of an observer. 
If the observer is moving in the gravitational field without any other forces except gravity, 
$u^{\mu}$  obeys the geodesic equation:  
\begin{eqnarray}
     u_{\mu} &=& g_{\mu\nu} u^{\nu} = g_{\mu\nu} \frac{d x^{\nu}}{d\tau}\,, \\
    \frac{d u_{\mu}}{d\tau} &=& \frac{1}{2} g_{\alpha\beta , \mu} u^{\alpha} u^{\beta}\,,\label{eq:geodeu}\\
    u_{\mu} u^{\mu} &=& -1\,,
\end{eqnarray}
where $\tau$ is the proper time as an affine parameter along the geodesic. 

\subsection{The composition rule of 4-velocities}\label{sec:comporule}

Let us consider two observers $A$ and $B$, whose 4-velocities are $u_A^{\mu} \equiv u^{\mu}$ and $u_B^{\mu} \equiv \bar{u}^{\mu}$ respectively.  
They are at the same point in the spacetime and observer $B$ is moving from observer $A$ with relative velocity $V$.  
We can write the following composition rule \cite{elliseq1}: 
\begin{eqnarray}
    \bar{u}^{\mu} &=& \frac{u^{\mu} + V e^{\mu}}{\sqrt{1-V^2}}\,,\label{eq:compou}\\
    e_{\mu} e^{\mu} &=& 1\,, \\
    e_{\mu} u^{\mu} &=& 0\,, 
\end{eqnarray}
where the unit space-like vector $e^{\mu}$ represents the direction of observer $B$'s motion in the observer $A$'s rest frame. A simple proof of the composition rule is given in Appendix \ref{sec:AppeA}.  

The inverse relation of the composition rule, which is based on the observer $B$'s rest frame, is
\begin{eqnarray}
    {u}^{\mu} &=& \frac{\bar{u}^{\mu} - V \bar{e}^{\mu}}{\sqrt{1-V^2}}\,,\label{eq:invcompou}\\
    \bar{e}_{\mu} \bar{e}^{\mu} &=& 1\,, \\
    \bar{e}_{\mu} \bar{u}^{\mu} &=& 0\,, 
\end{eqnarray}
where $\bar{e}^{\mu}$ represents the direction of observer $A$'s motion in the observer $B$'s rest frame. 
Actually, Eq.~\eqref{eq:invcompou} shows that observer $A$ is moving in the direction $-\bar{e}^{\mu}$ with relative velocity $V$ in the observer $B$'s rest frame.  

Using Eqs.~\eqref{eq:compou} and \eqref{eq:invcompou} and eliminating $\bar{u}^{\mu}$, we obtain
\begin{eqnarray}
    \bar{e}^{\mu} &=& \frac{e^{\mu} + V u^{\mu}}{\sqrt{1-V^2}}\,. \label{eq:compoe}
\end{eqnarray}
In the same way, we can also obtain
\begin{eqnarray}
    {e}^{\mu} &=& \frac{\bar{e}^{\mu} - V \bar{u}^{\mu}}{\sqrt{1-V^2}}\,. \label{eq:invcompoe}
\end{eqnarray}

\subsection{The Lorentz factor}

From Eqs.~\eqref{eq:compou} and \eqref{eq:compoe}, we can directly calculate the Lorentz factor $\gamma$: 
\begin{equation}\label{eq:Lofac}
    \gamma \equiv \frac{1}{\sqrt{1-V^2}} = - u_{\mu} \bar{u}^{\mu} = e_{\mu} \bar{e}^{\mu}\,.
\end{equation}
The Lorentz factor $\gamma$ is the invariant 4-scalar because it is calculated from the inner product of the 4-vectors $u_{\mu}$ and 
$\bar{u}^{\mu}$, or $e_{\mu}$ and $\bar{e}^{\mu}$. 

\subsection{The decomposition of the propagation 4-vector of light}\label{sec:decolight}

We introduce the following decomposition of the propagation 4-vector $k^{\mu}$ of light with respect to the observer's 4-velocity.
Consider the observer $A$ with 4-velocity $u^{\mu}$.  Using $u^{\mu}$, $k^{\mu}$ is 
decomposed into \cite{ellis612}
\begin{eqnarray}
    k^{\mu} &=& \omega \left(u^{\mu} + \gamma^{\mu}\right)\,, \label{eq:kdecomp}\\
    \gamma_{\mu} \gamma^{\mu} &=& 1 \,, \\
    \gamma_{\mu} u^{\mu} &=& 0\,,
\end{eqnarray}
where 
\begin{equation}\label{eq:omega}
    \omega \equiv - k_{\mu} u^{\mu} 
\end{equation}
is the frequency measured by observer $A$, and the space-like unit vector $\gamma^{\mu}$ represents the direction of light in the observer $A$'s rest frame.  

The decomposition can also be made with respect to the observer $B$'s 4-velocity $\bar{u}^{\mu}$: 
\begin{eqnarray}
    k^{\mu} &=& \bar{\omega} \left(\bar{u}^{\mu} + \bar{\gamma}^{\mu}\right)\,, \label{eq:kdecompbar}\\
    \bar{\gamma}_{\mu} \bar{\gamma}^{\mu} &=& 1 \,, \\
    \bar{\gamma}_{\mu} \bar{u}^{\mu} &=& 0\,,
\end{eqnarray}
where 
\begin{equation}\label{eq:baromega}
    \bar{\omega} \equiv - k_{\mu} \bar{u}^{\mu}
\end{equation}
is the frequency of the same light denoted by 
$k^{\mu}$ measured by the moving observer $B$, and $\bar{\gamma}^{\mu}$ represents the direction of light in the observer $B$'s rest frame.  

\subsection{The redshift}\label{sec:red}

The redshift\cite{ellis610b}  $z$ of a source as measured by a ``standard'' observer 
with 4-velocity $u^{\mu}$ is defined by the ratio of the two frequencies at the different points in the spacetime: 
\begin{equation}\label{eq:redshift}
    1 + z  \equiv \frac{\omega_{\rm source}}{\omega_{\rm observer}} 
    = \frac{(k_{\mu} u^{\mu})_{\rm source}}{(k_{\mu} u^{\mu})_{\rm observer}}\ .
\end{equation}
If the light, denoted by the tangent vector $k_{\mu}$, is propagating through the static spacetime, say, the Schwarzschild, $z$ is called the gravitational redshift.  (In case $z < 0$, it is called the blueshift.)   
If the light is propagating through the expanding universe, the same Eq.~(\ref{eq:redshift}) is called the cosmological redshift.

\subsection{The Doppler effect}\label{sec:Dop}

Whereas the ``standard'' redshift is defined by Eq.~(\ref{eq:redshift}), the Doppler effect is expressed by the following ratio of the two frequencies at the same point in the spacetime: 
\begin{equation}
    \frac{\bar{\omega}_{\rm source}}{\omega_{\rm source}} = 
    \frac{(k_{\mu} \bar{u}^{\mu})_{\rm source}}{(k_{\mu} u^{\mu})_{\rm source}}
        \quad\mbox{or}\quad
    \frac{\omega_{\rm observer}}{\bar{\omega}_{\rm observer}} = 
    \frac{(k_{\mu} u^{\mu})_{\rm observer}}{(k_{\mu} \bar{u}^{\mu})_{\rm observer}} \, ,
\end{equation}
where $\bar{u}^{\mu}$, as defined by Eq.~(\ref{eq:compou}), is the 4-velocity of the moving observer or source with relative velocity $V$. 
As Narlikar\cite{nar} wrote, ``in general relativity one cannot talk of a velocity of relative motion between two objects separated spatially''.  Then, he proceeded to the parallel transport argument for the cosmological and Doppler shifts.  
Compared to the Narlikar's argument, our definition of the Doppler effect is much more simplified. 
It is guaranteed to hold in general relativity by definition.  

In general, therefore, the redshift $\bar{z}$ of a moving source by a moving observer can be decomposed into the ``standard'' redshift part and the Doppler terms as follows: 
\begin{eqnarray}
    1 + \bar{z} &\equiv& \frac{\bar{\omega}_{\rm source}}{\bar{\omega}_{\rm observer}} \\
     &=& \frac{\bar{\omega}_{\rm source}}{{\omega}_{\rm source}}\cdot (1 + z) \cdot
    \frac{{\omega}_{\rm observer}}{\bar{\omega}_{\rm observer}}\, . \label{eq:barz}
\end{eqnarray}
In special relativity, there is neither gravitational nor cosmological redshifts, then $z=0$.  Therefore, from Eq.~\eqref{eq:barz}, we recover the usual Doppler formula
\begin{eqnarray}
    \frac{\bar{\omega}_{\rm source}}{\bar{\omega}_{\rm observer}} 
     &=& \frac{\bar{\omega}_{\rm source}}{{\omega}_{\rm source}}\cdot
    \frac{{\omega}_{\rm observer}}{\bar{\omega}_{\rm observer}} \quad\mbox{in special relativity.}
\end{eqnarray}

The relation between $\omega$ of Eq.~\eqref{eq:omega} and 
$\bar{\omega}$ of Eq.~\eqref{eq:baromega} can be obtained in the following way.  
Using Eqs.~\eqref{eq:invcompou} and \eqref{eq:kdecompbar}, 
\begin{eqnarray}
    \omega &=& - k_{\mu} u^{\mu} \\
    &=& - \bar{\omega} \left(\bar{u}_{\mu} + \bar{\gamma}_{\mu}\right)
    \frac{\bar{u}^{\mu} - V \bar{e}^{\mu}}{\sqrt{1-V^2}} \\
    &=& \bar{\omega} \frac{1 + V \cos\bar{\theta}}{\sqrt{1-V^2}}\,,
\end{eqnarray}
where $\cos\bar{\theta} \equiv \bar{\gamma}_{\mu}\bar{e}^{\mu}$, namely,  
$\bar{\theta}$ represents the angle between the direction of motion and the direction of light propagation in observer $B$'s rest frame. 

The Doppler formula for the moving observer is then
\begin{equation} \label{eq:DopB}
    \bar{\omega} = \omega \frac{\sqrt{1-V^2}}{1 - V \cos\bar{\vartheta}}\,,
\end{equation}
where $\bar{\vartheta} \equiv \pi - \bar{\theta}$ is the angle of incidence in the observer $B$'s rest frame.  
Actually, $\bar{\vartheta}$ is the angle between $\bar{e}^{\mu}$ and the direction of the source 
$- \bar{\gamma}^{\mu}$. In particular for the observer moving away from the source at rest, setting $\bar{\vartheta} = \pi$ in Eq.~\eqref{eq:DopB} and inserting it to Eq.~\eqref{eq:barz}, we obtain
\begin{equation}\label{eq:37}
    1 + \bar{z} = (1+z)\cdot\frac{{\omega}_{\rm observer}}{\bar{\omega}_{\rm observer}} = 
    (1+z)\cdot\sqrt{\frac{1+V}{1-V}}\  .
\end{equation}

The formula for the moving source can also be obtained in the following way.  
Using Eqs.~\eqref{eq:compou} and \eqref{eq:kdecomp}, 
\begin{eqnarray}
    \bar{\omega} &=& - k_{\mu} \bar{u}^{\mu} \\
    &=& - \omega \left(u^{\mu} + \gamma^{\mu}\right)\frac{u^{\mu} + V e^{\mu}}{\sqrt{1-V^2}} \\
    &=& \omega \frac{1 - V \cos{\theta}}{\sqrt{1-V^2}}\,,
\end{eqnarray}
where $\cos{\theta} \equiv {\gamma}_{\mu}{e}^{\mu}$ and ${\theta}$ represents the angle between the direction of motion and the direction of light propagation in observer $A$'s rest frame. 
The Doppler formula for the moving source is then
\begin{equation} \label{eq:DopA}
    \omega = \bar{\omega} \frac{\sqrt{1-V^2}}{1 + V \cos{\vartheta}}\,,
\end{equation}
where $\vartheta \equiv \pi - \theta$ is the angle of incidence in observer $A$'s rest frame.  
In particular for the source moving away from the observer at rest, setting $\vartheta = 0$ in Eq.~\eqref{eq:DopA} and  and inserting it to Eq.~\eqref{eq:barz}, we obtain
\begin{equation}
    1 + \bar{z} = \frac{\bar{\omega}_{\rm source}}{{\omega}_{\rm source}} \cdot (1+z) =
    (1+z)\cdot\sqrt{\frac{1+V}{1-V}}\ ,
\end{equation}
which gives the same result as of Eq.~\eqref{eq:37}.  

\subsection{The aberration}\label{sec:aberration}

Using Eqs.~\eqref{eq:DopB} and \eqref{eq:DopA} and eliminating $\bar{\omega}/\omega$, we obtain the formula for the aberration of light:
\begin{equation}\label{eq:aberration}
    \cos\bar{\vartheta} = \frac{\cos\vartheta + V}{1 + V \cos\vartheta}\,.
\end{equation}
A useful inequality for $0 < \vartheta < \pi$, $ 0 < \bar{\vartheta} < \pi$ is 
\begin{eqnarray}
    \cos\bar{\vartheta} - \cos{\vartheta} &=& \frac{V \sin^2\vartheta}{1 +V\cos\vartheta} > 0 \,,\\
    \therefore\ \ \bar{\vartheta} &<& {\vartheta} \,.
\end{eqnarray}

The formulae Eqs.~\eqref{eq:DopB}, \eqref{eq:DopA}, and \eqref{eq:aberration} look quite well known and seem nothing new. 
However, 
we emphasize that all quantities $\omega, \bar{\omega}, \vartheta, \bar{\vartheta}$, and $V$ are defined as the invariant 4-scalars.  
The formulae hold in any coordinate systems in any spacetime, in general relativity as well as in special relativity.

\section{The gravitational redshift and the Doppler effect in the Schwarzschild spacetime}\label{sec:Schwa}

\subsection{The metric of the Schwarzschild spacetime}

The metric of the Schwarzschild spacetime is
\begin{equation}\label{eq:Sch}
    ds^2 = -\left(1-\frac{r_g}{r}\right)  dt^2 + \frac{dr^2} {1-\frac{r_g}{r}} + r^2(d\theta^2 + \sin^2\theta \,d\phi^2)\,, 
\end{equation}
where $r_g \equiv 2 G M$ is the Schwarzschild radius.

\subsection{The solution of the null geodesic equation}

Thanks to the spherically symmetric property of the Schwarzschild spacetime, we may consider the orbit of light to be confined to the equatorial plane. Then, 
\begin{eqnarray}
    x^2 &=& \theta = \frac{\pi}{2}\,, \\
    k^2 &=& \frac{d x^2}{dv} = 0\,.
\end{eqnarray}
The Schwarzschild metric Eq.~\eqref{eq:Sch} does not depend on $x^0 = t, \, x^3 = \phi$, then the following conserved quantities are immediately obtained from Eq.~\eqref{eq:geodek}.  
\begin{eqnarray}
    k_0 &=& \mbox{const.} \equiv - \omega_c\,, 
    \ \ \therefore\ \ k^0 = \frac{\omega_c}{1 - \frac{r_g}{r}}\,, \\
    k_3 &=& \mbox{const.} \equiv l \,, \ \ 
    \therefore\ \ k^3 = \frac{l}{r^2}\,.
\end{eqnarray}
Finally, the null condition $k_{\mu} k^{\mu} = 0$ provides the equation for $k^1$ as follows:
\begin{equation}
    \left(k^1\right)^2 = \left(\frac{dr}{dv} \right)^2 
    = \omega_c^2 - \left(1-\frac{r_g}{r} \right) \frac{l^2}{r^2}\,.
\end{equation}

Particularly for radially propagating light, setting $l = 0$ yields
\begin{equation} \label{eq:radprok}
    k_{\mu} = (k_0, k_1, 0, 0) 
    = \left(-\omega_c, \pm \frac{\omega_c}{1-\frac{r_g}{r}}, 0, 0 \right)\,.
\end{equation}

\subsection{The 4-velocity of the observer at rest}

If an observer is at rest in the Schwarzschild spacetime, the 4-velocity is written as
\begin{equation} \label{eq:urest}
    u^{\mu} = \left(\frac{1}{\sqrt{1-\frac{r_g}{r}}}, 0, 0, 0\right) \,.
\end{equation}

\subsection{The 4-velocity of the moving observer on the geodesic}

In order to distinguish a moving observer from an observer at rest, 
we express the 4-velocity of the moving observer with a bar ($\bar{\ }$), as $\bar{u}^{\mu}$.  
If the observer is moving in the gravitational field without any other force except gravity, 
it obeys the geodesic equation Eq.~\eqref{eq:geodeu}.  
Again, we may consider the trajectory to be confined to the equatorial plane. Then, 
\begin{eqnarray}
    x^2 &=& \theta = \frac{\pi}{2}\,, \\
    \bar{u}^2 &=& \frac{d x^2}{d\tau} = 0\,,
\end{eqnarray}
and the following conserved quantities are immediately obtained from Eq.~\eqref{eq:geodeu}
\begin{eqnarray}
    \bar{u}_0 &=& \mbox{const.} \equiv - \epsilon\,, 
    \ \ \therefore\ \ \bar{u}^0 = \frac{\epsilon}{1 - \frac{r_g}{r}}\,, \\
    \bar{u}_3 &=& \mbox{const.} \equiv \ell \,, \ \ 
    \therefore\ \ \bar{u}^3 = \frac{\ell}{r^2}\,.
\end{eqnarray}
The condition $\bar{u}_{\mu} \bar{u}^{\mu} = -1$ provides the equation for 
$\bar{u}^1$ as
\begin{equation} 
    \left(\bar{u}^1\right)^2 = \left(\frac{dr}{d\tau} \right)^2 
    = \epsilon^2 - \left(1-\frac{r_g}{r} \right) \left(1 +\frac{\ell^2}{r^2}\right)\,. \label{eq:u1}
\end{equation}

\subsubsection{The 4-velocity of the moving observer on the radial geodesic}

If the observer is moving in radial direction, $\ell = 0$. Then, Eq.~\eqref{eq:u1} is
\begin{eqnarray}
    \left(\bar{u}^1\right)^2  
    = \epsilon^2 - \left(1-\frac{r_g}{r} \right)\,. 
\end{eqnarray}
As the initial condition, we impose $ \bar{u}^1 =  0$ at $r = r_i$. 
Then, 
\begin{equation}
    \epsilon = \sqrt{1 - \frac{r_g}{r_i}}\,.
\end{equation}
Therefore, the 4-velocity $\bar{u}^{\mu}$ of the moving observer on the radial geodesic is
\begin{equation}\label{eq:urad}
    \bar{u}^{\mu} = \left(\bar{u}^0, \bar{u}^1, 0, 0 \right) 
    = \left(\frac{ \sqrt{1 - \frac{r_g}{r_i}}}{1 - \frac{r_g}{r}}, 
            -\sqrt{\frac{r_g}{r} - \frac{r_g}{r_i}}, 0, 0\right)\,.
\end{equation}

For later use, we calculate the Lorentz factor of the radial geodesic motion:
\begin{equation}\label{eq:radLofac}
    \gamma \equiv \frac{1}{\sqrt{1-V^2}} 
    = - u_{\mu}(r) \bar{u}^{\mu}(r) 
    = \sqrt{\frac{ 1 - \frac{r_g}{r_i}}{1 - \frac{r_g}{r}}} \,.
\end{equation}

\subsubsection{The 4-velocity of the moving observer on the circular geodesic}

For circular motion, $r = \mbox{const.}$, Eq.~\eqref{eq:u1} reads 
\begin{equation} \label{eq:e2}
    \epsilon^2 = \left(1-\frac{r_g}{r} \right) \left(1 +\frac{\ell^2}{r^2}\right) \,.
\end{equation}
Differentiating it with $r$, we obtain
\begin{equation} \label{eq:de2}
    \frac{r_g}{r} - \frac{\ell^2}{r^2} \left( 2 - 3 \frac{r_g}{r}\right) = 0\,.
 \end{equation}
Solving the simultaneous equations Eqs.~\eqref{eq:e2} and \eqref{eq:de2}, we obtain 
\begin{eqnarray}
    \frac{\ell}{r} &=& 
    \sqrt{\frac{\frac{1}{2}\frac{r_g}{r}}{1 - \frac{3}{2}\frac{r_g}{r}}}\,, \\
    \epsilon &=& \frac{1-\frac{r_g}{r}}{\sqrt{1 - \frac{3}{2}\frac{r_g}{r}}}\,.
\end{eqnarray}
Therefore, the 4-velocity $\bar{u}^{\mu}$ of the moving observer on the circular geodesic is
\begin{equation}\label{eq:ucir}
    \bar{u}^{\mu} = (\bar{u}^0, 0, 0, \bar{u}^3) = \left(\frac{1}{\sqrt{1 - \frac{3}{2}\frac{r_g}{r}}}, 0, 0,
 \frac{1}{r}\frac{\sqrt{\frac{1}{2}\frac{r_g}{r}}}{\sqrt{1 - \frac{3}{2}\frac{r_g}{r}}}\right) \,.
\end{equation}

For later use, we calculate the Lorentz factor of the circular geodesic motion:
\begin{eqnarray}
    \gamma \equiv \frac{1}{\sqrt{1-V^2}} 
    &=& - u_{\mu}(r) \bar{u}^{\mu}(r) \\
    &=& \frac{\sqrt{1 - \frac{r_g}{r}}}{\sqrt{1-\frac{3}{2}\frac{r_g}{r}}}\,.
    \label{eq:cirLofac}
\end{eqnarray}

\subsubsection{The 4-velocity of the moving observer on the non-circular geodesic}
For non-circular motion, the geodesic equation cannot be solved analytically.  As long as the geodesic motion is on the bound orbit, however, there must be the maximum and minimum values of $r$, $r_{\rm max}$ and $r_{\rm min}$ at the points of $\displaystyle \frac{dr}{d\tau}=0$. Then, from Eq.~\eqref{eq:u1}, 
\begin{eqnarray}
    \epsilon^2 - \left( 1-\frac{r_g}{r_{\rm max}}\right) \left( 1 + \frac{\ell^2}{r_{\rm max}^2}\right) &=& 0\,, \\ 
    \epsilon^2 - \left( 1-\frac{r_g}{r_{\rm min}}\right) \left( 1 + \frac{\ell^2}{r_{\rm min}^2}\right)  &=& 0\,.
\end{eqnarray}
For later use, we solve these equations with respect to $\epsilon$ up to the linear order of $r_g$: 
\begin{equation}\label{eq:epeli}
    \epsilon \simeq 1 - \frac{r_g}{2(r_{\rm max} + r_{\rm min})} = 1 - \frac{r_g}{4a}\,,
\end{equation}
where
\begin{equation}
    a \equiv \frac{r_{\rm max} + r_{\rm min}}{2}
\end{equation}
corresponds to the semi-major axis in the Newtonian theory. 

\subsection{The gravitational redshift (or blueshift)}

Consider a light source at $r=r_1$ and an observer at $r=r_2$.  
Both source and observer are at rest.  
The source emits light denoted by $k_{\mu}$ at $r_1$, and it is received by the observer at rest at $r=r_2$. 
Using Eqs.~\eqref{eq:radprok} and \eqref{eq:urest}, the frequency at the source is
\begin{equation}
    \omega_1 = -k_{\mu}(r_1) u^{\mu}(r_1) = \frac{\omega_c}{\sqrt{1 - \frac{r_g}{r_1}}}\,.
\end{equation}
The frequency of the same light denoted by $k_{\mu}$, received at $r=r_2$ by the observer at rest, is
\begin{equation}
    \omega_2 = -k_{\mu}(r_2) u^{\mu}(r_2) = \frac{\omega_c}{\sqrt{1 - \frac{r_g}{r_2}}}\,.
\end{equation}
The ratio is
\begin{equation}\label{eq:gravred}
    \frac{\omega_2}{\omega_1} 
    = \frac{\sqrt{1 - \frac{r_g}{r_1}}}{\sqrt{1 - \frac{r_g}{r_2}}} 
    \ \ \left\{
\begin{array}{ll}
< 1& (\mbox{for}\ r_g < r_1 < r_2)\\ \ \\
> 1 & (\mbox{for}\ r_g < r_2 < r_1)
\end{array}
\right.\,.
\end{equation}
Both source and observer are at rest, and there is no relative motion. Still, the observed frequency at $r = r_2$ is different from that of the source at $r = r_1$.  
For $r_g < r_1 < r_2$, the observed frequency is smaller and it is called the gravitational redshift.   
On the other hand, for $r_g < r_2 < r_1$, it is called the gravitational blueshift.

\subsection{The Doppler effect due to moving observer on the radial geodesic}\label{sec:36}

Here we consider the case of the observer approaching to the source.  
Consider a light source at rest at $r=r_1$. 
The source emits light radially outward denoted by $k_{\mu}$ at $r_1$, and it is received by the observer in radial geodesic motion with $\bar{u}^{\mu}$ at $r=r_2 > r_1$. 

From Eq.~\eqref{eq:radprok}, $k_{\mu}$ for outgoing light is
\begin{equation}\label{eq:kout}
    k_{\mu} 
    = \left(-\omega_c, + \frac{\omega_c}{1-\frac{r_g}{r}}, 0, 0 \right)\,.
\end{equation}
Using Eq.~\eqref{eq:urad}, 
the frequency $\bar{\omega}_2$ observed by the moving observer at $r=r_2$ is
\begin{eqnarray}
    \bar{\omega}_2 &=& - k_{\mu}(r_2) \bar{u}^{\mu}(r_2) \\  
    &=& \frac{\omega_c}{1 - \frac{r_g}{r_2}} \left(\sqrt{1 - \frac{r_g}{r_i}} + \sqrt{\frac{r_g}{r_2} - \frac{r_g}{r_i}}\right) \,,
\end{eqnarray}
where we have assumed $r_i > r_2$. 

The ratio of the source and the observed frequencies can be divided into two parts:
\begin{eqnarray}
    \frac{\bar{\omega}_2}{\omega_1} 
    &=& \frac{\omega_2}{\omega_1}\cdot\frac{\bar{\omega}_2}{\omega_2} \\
    &=& \frac{\sqrt{1 - \frac{r_g}{r_1}}}{\sqrt{1 - \frac{r_g}{r_2}}} \cdot
    \frac{1}{\sqrt{1 - \frac{r_g}{r_2}}} \left(\sqrt{1 - \frac{r_g}{r_i}} + \sqrt{\frac{r_g}{r_2} - \frac{r_g}{r_i}}\right) \label{eq:69}\\
    &>& \frac{\sqrt{1 - \frac{r_g}{r_1}}}{\sqrt{1 - \frac{r_g}{r_2}}}\,,
\end{eqnarray}
where the first part is the gravitational redshift, and the second part denotes the Doppler effect due to the observer's motion towards the source. 





\subsection{The Doppler effect due to the moving source on the radial geodesic}\label{sec:37}

Here we consider the case of the source moving away from the observer.  
Consider two sources. One is at rest at $r=r_1$ and the emitted light is denoted by $k_{\mu}$ of \eqref{eq:kout}.  
The other is moving with 4-velocity of Eq.~\eqref{eq:urad} away from the observer, and emits light outward which is denoted by $k'_{\mu}$ as follows:
\begin{equation}\label{eq:kpout}
    k'_{\mu} 
    = \left(-\omega'_c, + \frac{\omega'_c}{1-\frac{r_g}{r}}, 0, 0 \right)\,.
\end{equation}
The frequencies observed in the source rest frames, respectively at $r=r_1$, are
\begin{eqnarray}
    \omega_1 &=& - k_{\mu}(r_1) u^{\mu}(r_1) = \frac{\omega_c}{\sqrt{1 - \frac{r_g}{r_1}}}\,, \\
    \bar{\omega}'_1 &=& - k'_{\mu}(r_1) \bar{u}^{\mu}(r_1) 
    = \frac{\omega'_c}{1 - \frac{r_g}{r_1}} \left(\sqrt{1 - \frac{r_g}{r_i}} + \sqrt{\frac{r_g}{r_1} - \frac{r_g}{r_i}}\right) \,.
\end{eqnarray}
We assume the two frequencies are the same at $r=r_1$. 
Then,  
\begin{eqnarray}
    \omega_1 &\equiv& \bar{\omega}'_1\,, \\
    \therefore \ \ \omega'_c &=& 
     \frac{\omega_c}{\sqrt{1 - \frac{r_g}{r_1}}} \left(\sqrt{1 - \frac{r_g}{r_i}} - \sqrt{\frac{r_g}{r_1} - \frac{r_g}{r_i}}\right)\,.
\end{eqnarray}

The sources emit lights radially outward, and are received by the observer at rest at $r=r_2 > r_1$. The observed frequencies at $r=r_2$ are 
\begin{eqnarray}
    \omega_2 &=& - k_{\mu}(r_2)  u^{\mu}(r_2) 
    = \frac{\omega_c}{\sqrt{1 - \frac{r_g}{r_2}}}\,,  \\
    \omega'_2 &=& - k'_{\mu}(r_2)  u^{\mu}(r_2) 
    = \frac{\omega'_c}{\sqrt{1 - \frac{r_g}{r_2}}}\,.  
\end{eqnarray}

The ratio of the source and the observed frequencies of the light emitted from the moving source is 
\begin{eqnarray}
    \frac{\omega'_2}{\bar{\omega}'_1} &=& \frac{\omega'_2}{\omega_1} \\
    &=& \frac{\omega_2}{\omega_1} \cdot\frac{\omega'_2}{\omega_2} 
    =\frac{\omega_2}{\omega_1} \cdot\frac{\omega'_c}{\omega_c} \\
    &=& \frac{\sqrt{1 - \frac{r_g}{r_1}}}{\sqrt{1 - \frac{r_g}{r_2}}} 
    \cdot \frac{1}{\sqrt{1 - \frac{r_g}{r_1}}} \left(\sqrt{1 - \frac{r_g}{r_i}} - \sqrt{\frac{r_g}{r_1} - \frac{r_g}{r_i}}\right) \label{eq:97}\\   
    &<& \frac{\sqrt{1 - \frac{r_g}{r_1}}}{\sqrt{1 - \frac{r_g}{r_2}}} \,.
\end{eqnarray}
The ratio is divided into two parts: the first part is the gravitational redshift, and the second part is the Doppler effect due to the moving source away from the observer.  

\subsection{The transverse Doppler effect due to moving source on the circular geodesic}\label{sec:38}

Let us consider two light sources.  
One is at rest at $r=r_1$ and emits light radially with $k_{\mu}$ of Eq.~\eqref{eq:kout}.  
The other is in circular geodesic motion of radius $r_1$ with 4-velocity $\bar{u}^{\mu}$ of Eq.~\eqref{eq:ucir}, and it also emits light with $\bar{k}_{\mu}$ of Eq.~\eqref{eq:kpout}.  
The frequencies at $r=r_1$ in the source rest frames are
\begin{eqnarray}
    \omega_1 &=& - k_{\mu}(r_1)  u^{\mu}(r_1) 
    = \frac{\omega_c}{\sqrt{1-\frac{r_g}{r_1}}} \,,\\
    \bar{\omega}'_1 &=& - k'_{\mu}(r_1)  \bar{u}^{\mu}(r_1) 
    =\frac{\omega'_c}{\sqrt{1-\frac{3}{2}\frac{r_g}{r_1}}} \,. 
\end{eqnarray}
We assume the two frequencies are the same at $r=r_1$. Then, 
\begin{eqnarray}
    \omega_1 &\equiv& \bar{\omega}'_1\,, \\
    \therefore \ \ \omega'_c &=& 
    \omega_c\frac{\sqrt{1-\frac{3}{2}\frac{r_g}{r_1}}}{\sqrt{1 - \frac{r_g}{r_1}}}\,.
\end{eqnarray}

The observer at rest at $r = r_2$ receives the lights. 
For the sake of simplicity, here we only consider the case $r_1 < r_2$. 
The observed frequencies are
\begin{eqnarray}
    \omega_2 &=& - k_{\mu} u^{\mu}(r_2) 
    = \frac{\omega_c}{\sqrt{1-\frac{r_g}{r_2}}} \,,\\
    \omega'_2 &=& - k'_{\mu} u^{\mu}(r_2) 
    = \frac{\omega'_c}{\sqrt{1-\frac{r_g}{r_2}}} \,.  
\end{eqnarray}

The ratio of the frequencies of light emitted from the source in circular geodesic motion is 
\begin{eqnarray}
    \frac{\omega'_2}{\bar{\omega}'_1} &=& \frac{\omega'_2}{\omega_1} \\
    &=& \frac{\omega_2}{\omega_1}\cdot \frac{\omega'_2}{\omega_2} 
    = \frac{\omega_2}{\omega_1}\cdot \frac{\omega'_c}{\omega_c}\\
    &=& \frac{\sqrt{1 - \frac{r_g}{r_1}}}{\sqrt{1 - \frac{r_g}{r_2}}} 
    \cdot \frac{\sqrt{1-\frac{3}{2}\frac{r_g}{r_1}}}{\sqrt{1 - \frac{r_g}{r_1}}} \label{eq:107}\\
    &<& \frac{\sqrt{1 - \frac{r_g}{r_1}}}{\sqrt{1 - \frac{r_g}{r_2}}}\,.
\end{eqnarray}
The ratio is divided into two parts.  The first part is the gravitational redshift as usual.  
For the observer at rest, the direction of the source's circular motion $e^{\mu}$ is perpendicular to the radial direction of the emitted light $\gamma^{\mu}$, hence $\cos\vartheta = 0$.  
Therefore, the second part can be regarded as the transverse Doppler effect due to the moving source.

\subsection{Compatibility with the Doppler formula}

We have already derived the general formula for the Doppler effect in Sec.~\ref{sec:Dop} for the relative velocity $V$ between the source and the observer.  
On the other hand, we have used the solutions of the geodesic equation in this section.  Then the frequencies are expressed as the functions of $r$, and $V$ does not appear explicitly.  
Here we examine the compatibility with the Doppler formula and the results obtained in the above Sec.~\ref{sec:36}-\ref{sec:38}.  

\subsubsection{The cases of the longitudinal Doppler effect}

The Doppler term for the the observer approaching radially to the source at $r=r_2$ is derived in Eq.~\eqref{eq:69}:
\begin{equation}\label{eq:109}
    \frac{\bar{\omega}_2}{\omega_2} = 
    \frac{1}{\sqrt{1 - \frac{r_g}{r_2}}} \left(\sqrt{1 - \frac{r_g}{r_i}} + \sqrt{\frac{r_g}{r_2} - \frac{r_g}{r_i}}\right)\,.
\end{equation}
Using the Doppler formula Eq.~\eqref{eq:DopB} and setting $\bar{\vartheta}=0$, the Doppler effect is
\begin{equation}\label{eq:110}
    \frac{\bar{\omega}}{\omega} = \sqrt{\frac{1+V}{1-V}}\,.
\end{equation}
The explicit form of $V$ in terms of $r_2$ can be obtained from the definition of the Lorentz factor Eq.~\eqref{eq:radLofac}.  
\begin{eqnarray}
    \gamma \equiv \frac{1}{\sqrt{1-V^2}} 
    &=& - u_{\mu}(r_2) \bar{u}^{\mu}(r_2) \\
    &=& \sqrt{\frac{ 1 - \frac{r_g}{r_i}}{1 - \frac{r_g}{r_2}}} \,, \\
    \therefore\ \ 
    V &=& \sqrt{\frac{\frac{r_g}{r_2} - \frac{r_g}{r_i}}{1 - \frac{r_g}{r_i}}}\,.
    \label{eq:Vlongi}
\end{eqnarray}
Inserting Eq.~\eqref{eq:Vlongi} into Eq.~\eqref{eq:110}, we can obtain the right-hand-side of Eq.~\eqref{eq:109}.  Therefore, the longitudinal Doppler effect Eq.~\eqref{eq:109} is compatible with the Doppler formula Eq.~\eqref{eq:110}. 


\subsubsection{The case of the transverse Doppler effect}

The transverse Doppler effect due to the moving source is derived in Eq.~\eqref{eq:107}:
\begin{equation}\label{eq:116}
    \frac{\omega'_c}{\omega_c} = \frac{\sqrt{1-\frac{3}{2}\frac{r_g}{r_1}}}{\sqrt{1 - \frac{r_g}{r_1}}}\,. 
\end{equation}
Using the Doppler formula Eq.~\eqref{eq:DopA} and setting $\vartheta = \pi/2$, the transverse Doppler effect is
\begin{equation}\label{eq:117}
    \frac{\omega}{\bar{\omega}} = \sqrt{1-V^2}\,.
\end{equation}
The explicit form of $V$ in terms of $r_1$ can be obtained from the definition of the Lorentz factor Eq.~\eqref{eq:cirLofac}.  
\begin{eqnarray}
    \gamma \equiv \frac{1}{\sqrt{1-V^2}} 
    &=& - u_{\mu}(r_1) \bar{u}^{\mu}(r_1) \\
    &=& \frac{\sqrt{1 - \frac{r_g}{r_1}}}{\sqrt{1-\frac{3}{2}\frac{r_g}{r_1}}}\,.
    \label{eq:Vtrans}
\end{eqnarray}
Inserting Eq.~\eqref{eq:Vtrans} into Eq.~\eqref{eq:117}, it is apparent that 
the transverse Doppler effect Eq.~\eqref{eq:116} is compatible with the Doppler formula Eq.~\eqref{eq:117}.

\section{The time dilation in the Schwarzschild spacetime}\label{sec:timedil}

\subsection{The gravitational time dilation}

Consider two clocks at rest at different positions in the gravitational field.  
The elapsed time $\varDelta T$ of clock is defined to be inversely proportional to the frequency $\omega$ of a particular light or electromagnetic wave observed in the clock's rest frame:   
\begin{equation}
    \varDelta T \propto \frac{1}{\omega}\,.
\end{equation}
Then, the ratio of the elapsed times of the clocks at rest at $r_1$, and $r_2$ is
\begin{equation}\label{eq:graT}
    \frac{\varDelta T_1}{\varDelta T_2} = \frac{\omega_2}{\omega_1} 
    = \frac{\sqrt{1 - \frac{r_g}{r_1}}}{\sqrt{1 - \frac{r_g}{r_2}}}  \,.
\end{equation}
Therefore, $\varDelta T_1 < \varDelta T_2$ for $r_g < r_1 < r_2$. 
Clocks in the strong gravitational gravitational field (i.e., near the strong gravitational source) tick slowly.  
This is called the gravitational time dilation.  

\subsection{The kinetic time dilation}

Consider two clocks at the same point in the gravitational field. 
Assume that clock $A$ is at rest with $u^{\mu}$ 
and ticks $\varDelta T$, and 
the other clock $B$ is moving 
with $\bar{u}^{\mu}$ of Eq.~\eqref{eq:compou} relative to clock $A$, 
ticking $\varDelta \bar{T}$.

As is explained in Appendix~\ref{sec:B2}, the ratio of the elapsed times is
\begin{equation}\label{eq:kinT}
    \frac{\varDelta \bar{T}}{\varDelta T} = \sqrt{1-V^2} 
    = \frac{1}{- u_{\mu} \bar{u}^{\mu}}\,.
\end{equation}
$\varDelta \bar{T} < \varDelta T$.  
Clocks in motion tick slowly. 
This is called the kinetic time dilation .

\subsection{Time dilation of the moving clock on the radial orbit}

Assume that clock $A$ is at rest at $r = r_i$ and ticking $\varDelta T_i$, 
and clock $B$ is moving on the radial geodesic 
with the initial condition $u^1 = 0$ at $r=r_i$, 
ticking $\varDelta \bar{T}$ at $r$.
The ratio of the elapsed time is
\begin{eqnarray}
    \frac{\varDelta \bar{T}}{\varDelta T_i} &=& 
    \frac{\varDelta {T}}{\varDelta T_i}\cdot
    \frac{\varDelta \bar{T}}{\varDelta T} \\
    &=& \frac{\sqrt{1 - \frac{r_g}{r}}}{\sqrt{1 - \frac{r_g}{r_i}}}
    \cdot \frac{1}{- u_{\mu}(r) \bar{u}^{\mu}(r)} \\
    &=& \frac{1 - \frac{r_g}{r}}{1 - \frac{r_g}{r_i}}\,,
\end{eqnarray}
where we have used Eq.~\eqref{eq:radLofac}.

\subsection{Time dilation of the moving clock on the circular orbit}

Assume that clock $A$ is at rest at $r = r_1$ and ticking $\varDelta T_1$, 
and clock $B$ is moving on the circular orbit of radius $r$, 
ticking $\varDelta \bar{T}$. 
The ratio of the elapsed time is
\begin{eqnarray}
    \frac{\varDelta \bar{T}}{\varDelta T_1} &=& 
    \frac{\varDelta {T}}{\varDelta T_1} \frac{\varDelta \bar{T}}{\varDelta T} \\
    &=& \frac{\sqrt{1 - \frac{r_g}{r}}}{\sqrt{1 - \frac{r_g}{r_1}}}\cdot
    \frac{1}{- u_{\mu}(r) \bar{u}^{\mu}(r)} \\
    &=& \frac{\sqrt{1 - \frac{3}{2}\frac{r_g}{r}}}{\sqrt{1 - \frac{r_g}{r_1}}} \,, \label{eq:tdcirc}
\end{eqnarray}
where we have used Eq.~\eqref{eq:cirLofac}.  
The time dilation formula Eq.~\eqref{eq:tdcirc} looks quite well known and seems nothing new.  However, 
we would like to point out that this equation exactly holds without approximation in general relativity.  Just for reference, a conventional derivation based on the Newtonian analogy is given in Appendix~\ref{sec:B3}.

\subsection{Time dilation of the moving clock on the elliptical orbit}

Assume that clock $A$ is at rest at $r = r_1$ and ticking $\varDelta T_1$, 
and clock $B$ is moving on the non-circular bound orbit,
ticking $\varDelta \bar{T}$ at $r$ changing from time to time.   
The Lorentz factor is
\begin{equation}
    \gamma = - u_{\mu}(r) \bar{u}^{\mu}(r) = \frac{\epsilon}{\sqrt{1-\frac{r_g}{r}}}\,. 
\end{equation}
Then the ratio of the elapsed times is
\begin{eqnarray}
    \frac{\varDelta \bar{T}}{\varDelta T_1} &=& 
    \frac{\varDelta {T}}{\varDelta T_1} \frac{\varDelta \bar{T}}{\varDelta T} \\
    &=& \frac{\sqrt{1 - \frac{r_g}{r}}}{\sqrt{1 - \frac{r_g}{r_1}}}\cdot
    \frac{1}{- u_{\mu}(r) \bar{u}^{\mu}(r)} \\
    &=& \frac{1 - \frac{r_g}{r}}{\epsilon \sqrt{1 - \frac{r_g}{r_1}}}\,.
\end{eqnarray}

For non-circular motion, the geodesic equation cannot be solved analytically. Using Eq.~\eqref{eq:epeli}, the following expression is valid up to the linear order of $r_g$: 
\begin{equation}
    \frac{\varDelta \bar{T}}{\varDelta T_1} \simeq
    \frac{1}{\sqrt{1 - \frac{r_g}{r_1}}}\left(1 + \frac{r_g}{4 a}- \frac{r_g}{r}\right)\,.
\end{equation}

For non-circular bound orbits, $r$ changes from time to time.  
Up to the linear order of $r_g$, however, we can treat $r$ as the elliptical orbit.  Then, the time average of the elliptical orbit per cycle is
\begin{equation}
    \left\langle \frac{1}{r} \right\rangle = \frac{1}{a}\,, 
\end{equation}
where $a$ is the semi-major axis of the elliptical orbit.  
Using this result, the ratio of the elapsed times, after the time average per cycle, is
\begin{eqnarray}
    \frac{\left\langle \varDelta \bar{T}\right\rangle}{\varDelta T_1} 
    &\simeq&
    \frac{1}{\sqrt{1 - \frac{r_g}{r_1}}}\left(1 + \frac{r_g}{4 a}- \left\langle\frac{r_g}{r}\right\rangle\right) \\
    &\simeq& \frac{\sqrt{1 - \frac{3}{2} \frac{r_g}{a}}}{\sqrt{1 - \frac{r_g}{r_1}}} \,,\label{eq:tdnoncirc}
\end{eqnarray}
which is valid up to the linear order of $r_g$. 
The time dilation of the moving clock on elliptical orbit, after averaging per cycle, depends only on the semi-major axis, irrespective of the eccentricity.  
It is also quite impressive to compare this result Eq.~\eqref{eq:tdnoncirc} with Eq.~\eqref{eq:tdcirc}.  
Replacing the circular radius $r$ in Eq.~\eqref{eq:tdcirc} with the semi-major axis $a$ reproduces the result Eq.~\eqref{eq:tdnoncirc}.

\section{The cosmological redshift and the Doppler effect in the expanding universe}

\subsection{The metric of the expanding universe}

The Friedmann-Lema\^{i}tre-Robertson-Walker (FLRW) metric which describes the expanding universe is
\begin{equation}
    ds^2 = -a^2(\eta) d\eta^2 + a^2(\eta) 
    \Bigl(d\chi^2 + \sigma^2(\chi)\left(d\theta^2 + \sin^2\theta \,d\phi^2 \right) \Bigr) \,, \label{eq:FLRW}
\end{equation}
where we use the conformal time coordinate $\eta$, and
\begin{equation}
    \sigma(\chi) 
    =\left\{
        \begin{array}{ll}
        \frac{\sin \left( \sqrt{k} \chi\right)}{\sqrt{k}} & (k > 0) \,,\\ 
        \chi & (k = 0) \,, \\
        \frac{\sinh \left( \sqrt{|k|} \chi\right)}{\sqrt{|k|}} & (k < 0) \,,
        \end{array}
        \right.
\end{equation}
and $k$ is the curvature constant of the homogeneous and isotropic space.  

\subsection{The solution of the null geodesic equation}

Although the metric depends on $x^0 = \eta$, we can show $k_0$ is constant:  
\begin{eqnarray}
    \frac{d k_0}{dv} &=& \frac{1}{2} g_{\alpha\beta, 0} k^{\alpha} k^{\beta} 
    = \frac{1}{a}\frac{d a}{d\eta} g_{\alpha\beta} k^{\alpha} k^{\beta} = 0 \,, \\
    \therefore\ \ k_0 &=& \mbox{const.} \equiv - \omega_c \,,
\end{eqnarray}
where the null condition $g_{\alpha\beta} k^{\alpha} k^{\beta} = 0$ is used.  

Because of the homogeneous and isotropic property of the FLRW universe, 
it is sufficient to consider radially propagating null geodesics.
Then, $k^2 = k^3 = 0$, and the null condition provides the relation $k^1 = \pm k^0$.  
Finally, $k_{\mu}$ for the radially propagating light in the FLRW universe is
\begin{equation}\label{eq:55}
    k_{\mu} = (-\omega_c, \pm \omega_c, 0, 0)\,.
\end{equation}

\subsection{The 4-velocity of the comoving observer}

The spatial components of the comoving observer's 4-velocity vanish, $u^{i} = 0$ by the definition.  
Using the condition $u_{\mu} u^{\mu} = -1$, we obtain
\begin{equation}\label{eq:56}
    u^{\mu} = \left(\frac{1}{a}, 0, 0, 0 \right)\,.
\end{equation}

\subsection{The 4-velocity of the moving observer with peculiar velocity}

The 4-velocity of the moving observer $\bar{u}^{\mu}$ with peculiar velocity $V$ relative to the comoving observer is expressed by Eq.~\eqref{eq:compou} or Eq.~\eqref{eq:invcompou}.  

\subsection{The cosmological redshift}

The frequency of light $\omega$ observed by the comoving observer is, from Eqs.~\eqref{eq:55} and \eqref{eq:56},  
\begin{equation}
    \omega \equiv -k_{\mu} u^{\mu} = \frac{\omega_c}{a}\,,
\end{equation}
which depends only on the time  through the scale factor $a(\eta)$.  
The redshift $z$ 
Eq.~\eqref{eq:redshift} is 
\begin{equation}
    1 + z = \frac{\omega_{\rm source}}{\omega_{\rm observer}}  = \frac{a_{\rm observer}}{a_{\rm source}} > 1\,.
\end{equation} 
This is the well-known formula for the cosmological redshift.  

\subsection{The Doppler effect due to the moving observer}

The redshift $\bar{z}$ measured by the moving observer with $\bar{u}^{\mu}$ is, with the help of Eq.~\eqref{eq:DopB},  
\begin{eqnarray}
    1 + \bar{z} &=& \frac{\omega_{\rm source}}{\bar{\omega}_{\rm observer}} \\
    &=& \frac{\omega_{\rm source}}{{\omega}_{\rm observer}}\cdot\frac{\omega_{\rm observer}}{\bar{\omega}_{\rm observer}} \\
    &=& (1 + z)\cdot \frac{1 - V \cos\bar{\vartheta}}{\sqrt{1-V^2}}\,,\label{eq:cosDop}
\end{eqnarray}
where, as explained in Sec.~\ref{sec:Dop}, $\bar{\vartheta}$ is the angle of incidence in the moving observer's rest frame.  
The dipole anisotropy, which is proportional to $\cos\bar{\vartheta}$, naturally appears as the Doppler effect due to the observer's peculiar motion.  
Note that the amplitude for the longitudinal Doppler effect when $\bar{\vartheta} = 0$ is 
$V/\sqrt{1-V^2}$.  
We also observe the transverse Doppler effect
\begin{equation}
    1 + \bar{z} = (1 + z)\cdot \frac{1}{\sqrt{1-V^2}}
    > 1 + z \quad \mbox{for}\ \bar{\vartheta} = \frac{\pi}{2}\,. 
\end{equation}

\subsection{The anisotropy of the cosmic microwave background by the Doppler effect}

Let us express the temperature of the cosmic microwave background (CMB) at the recombination epoch as $T_{\gamma}$, and those observed by the observer at rest and by the moving observer as $T_{\gamma 0}$ and $\bar{T}_{\gamma 0}$, respectively.  Then, 
\begin{equation}
    1 + z = \frac{T_{\gamma}}{T_{\gamma 0}}, \quad 
    1 + \bar{z} = \frac{T_{\gamma}}{\bar{T}_{\gamma 0}}\,. 
\end{equation}
From Eq.~\eqref{eq:cosDop}, we obtain
\begin{equation}\label{eq:barT}
    \bar{T}_{\gamma 0} = T_{\gamma 0} \frac{\sqrt{1-V^2}}{1 - V \cos\bar{\vartheta}}\,,
\end{equation}
where $\bar{\vartheta}$ is the angle between the direction of the observer's velocity and the incoming CMB, in the moving observer's frame.  
(See also Weinberg (1972)\cite{wein1972} and Weinberg (2008)\cite{wein2008}.)    
The Taylor expansion of Eq.~\eqref{eq:barT} in powers of $V$ gives
\begin{equation}
    \delta T_{\gamma 0} \equiv \bar{T}_{\gamma 0} - T_{\gamma 0} 
    \simeq T_{\gamma 0} \left[V \cos\bar{\vartheta}  
    + \frac{V^2}{2} \left(2 \cos^2\bar{\vartheta} -1 \right) + \cdots \right]\,.
\end{equation}
Although the amplitudes are small (because $|V| \ll 1$), higher order terms naturally includes the kinematic quadrupole component caused by the Doppler effect.

\section{Conclusion}

We have presented a unified treatment of the gravitational and cosmological redshift, the Doppler effect due to the moving observer or source, and the time dilation in the gravitational field in the framework of general relativity.

We have applied it to the cases of moving observer or light source in the gravitational field, and obtained the Doppler effect formula with the velocity $V$ of the observer or the source, in addition to the standard gravitational or cosmological redshift.  
In particular, the longitudinal and the transverse Doppler effects have explicitly been given which hold for the moving observer or light source in the gravitational field described by the Schwarzschild metric.  

We have also examined the time dilation of the moving clock in the gravitational field.  

We have confirmed that the well-known formula for the ratio of the elapsed times
\begin{equation*}
    \frac{\Delta \bar{T}}{\Delta T_1} = \frac{\sqrt{1 - \frac{3}{2}\frac{r_g}{r}}}{\sqrt{1 - \frac{r_g}{r_1}}}\,, 
\end{equation*}
where  $\Delta \bar{T}$ is of the moving clock on circular orbit with radius $r$ and $\Delta T_1$ is of the observer at rest $r=r_1$,  
exactly holds without approximation.  

We have also derived the time dilation of the moving clock on the elliptical orbit with the semi-major axis $a$, based on the unified treatment.  
The ratio of the elapsed times, after the time average per cycle, is 
\begin{equation*}
    \frac{\langle\Delta \bar{T}\rangle} {\Delta T_1} \simeq 
    \frac{\sqrt{1 - \frac{3}{2}\frac{r_g}{a}}} {\sqrt{1 - \frac{r_g}{r_1}}}\,,  
\end{equation*}
which holds up to the first order of $r_g$.  

We have applied our unified treatment to the cosmological redshift and obtained the Doppler effect formulae which exactly hold in the general relativistic framework.  We have observed the existence of the transverse Doppler effect due to the observer's peculiar motion in the expanding universe.  

Needless to say, the unified treatment presented in this paper can also be applied to the special relativistic cases. It means that the special relativistic effects can also be understood without the Lorentz transformation, which are summarized in Appendix~\ref{sec:B} for reader's convenience.

\appendix

\section{A proof of the composition rule of 4-velocities}\label{sec:AppeA}

Consider two observers $A$ and $B$, whose 4-velocities are $u^{\mu}$ and $\bar{u}^{\mu}$ respectively, are at the same point $P$ in the spacetime and the observer $B$ moves away from the observer $A$ with relative velocity $V$. 
Under a suitable coordinate transformation, we can always take the local frame which is momentarily comoving with observer $A$ at the given point $P$.  In this coordinate system, 
\begin{equation}
    u^{\mu} = \frac{d x^{\mu}}{d\tau} = (1, 0, 0, 0)\,. 
\end{equation}
Assume that the observer $B$ moves in the $x$ direction with velocity $V$. Then, 
\begin{equation}
    \frac{d\bar{x}^i}{d\bar{t}} = \frac{\bar{u}^i}{\bar{u}^0} = (V, 0, 0)\,.
\end{equation}
From $\bar{u}_{\mu} \bar{u}^{\mu} = -1$, we obtain
\begin{eqnarray}
    \bar{u}^{\mu} &=& \left(\frac{1}{\sqrt{1-V^2}}, \frac{V}{\sqrt{1-V^2}}, 0, 0 \right) \\
    &=& \frac{1}{\sqrt{1-V^2}} \,(1, 0, 0, 0) + \frac{V}{\sqrt{1-V^2}}\,(0, 1, 0, 0) \\
    &=& \frac{1}{\sqrt{1-V^2}}\, u^{\mu} + \frac{V}{\sqrt{1-V^2}}\, e^{\mu}\,. \\
    \therefore\ \ \bar{u}^{\mu} &=& \frac{u^{\mu} + Ve^{\mu}}{\sqrt{1-V^2}}\,, \label{eq:A6}
\end{eqnarray}
where the unit space-like vector $e^{\mu} = (0, 1, 0, 0)$ represents the direction of motion of the observer $B$ in the observer $A$'s rest frame, and satisfies $e_{\mu} e^{\mu} = 1, \, e_{\mu} u^{\mu} = 0$.   
Since Eq.~\eqref{eq:A6} is the vector equation, it turns out to hold in any coordinate systems in any spacetime, in general relativity as well as in special relativity. 

\section{Understanding the special relativistic effects without the Lorentz transformation}\label{sec:B}

The basic principles and equations presented in this paper hold also in special relativistic situations.  All we have to do is just $g_{\mu\nu} \Rightarrow \eta_{\mu\nu}$. 
Using the results presented in this paper, we can understand special relativistic effects in the general relativistic framework, without using the Lorentz transformation.  
Just for readers' convenience, we gather related equations in this Appendix.  

\subsection{Observers' 4-velocities and the composition rule}

For the sake of simplicity, we state observer $A$ with 4-velocity $u^{\mu}$ is ``at rest'' and observer $B$ with 4-velocity $\bar{u}^{\mu}$ is ``moving'' with relative velocity $V$.  
The relation between $u^{\mu}$ and $\bar{u}^{\mu}$ is presented as the composition rule of 4-velocities in Sec.~\ref{sec:comporule}.  Here we gather the related equations Eqs.~\eqref{eq:compou}, \eqref{eq:invcompou}, \eqref{eq:compoe}, \eqref{eq:invcompoe}, and list again in the following:
\begin{eqnarray}
    \bar{u}^{\mu} &=& \frac{u^{\mu} + V e^{\mu}}{\sqrt{1-V^2}}\,, \label{eq:B1}\\
    \bar{e}^{\mu} &=& \frac{e^{\mu} + V u^{\mu}}{\sqrt{1-V^2}}\,, \label{eq:B2} \\
    {u}^{\mu} &=& \frac{\bar{u}^{\mu} - V \bar{e}^{\mu}}{\sqrt{1-V^2}}\,, \label{eq:B3}\\
    {e}^{\mu} &=& \frac{\bar{e}^{\mu} - V \bar{u}^{\mu}}{\sqrt{1-V^2}}\,,
\end{eqnarray}
where the unit space-like vector $e^{\mu}$ represents the direction of motion of the observer $B$ in the observer $A$'s rest frame, $\bar{e}^{\mu}$ represents the direction of motion of the observer $A$ in the observer $B$'s rest frame.  
Actually, Eq.~\eqref{eq:B3} shows that observer $A$ moves in the direction $-\bar{e}^{\mu}$ with relative velocity $V$ in the observer $B$'s rest frame.  

Note that these relations do not represent any coordinate transformations even if they somewhat look similar to the Lorentz transformation.  
They just represent that some 4-vectors are expressed by the linear combinations of other 4-vectors at the same point in the spacetime.  
Since they are vector equations, they hold in any coordinate systems in any spacetime, including the Minkowski spacetime in special relativity.   

\subsection{Time dilation}\label{sec:B2}

Consider a clock moving with 4-velocity $\bar{u}^{\mu}$.  The 4-vector 
\begin{equation}
    t^{\mu} \equiv \varDelta T_0 \,\bar{u}^{\mu}
\end{equation}
represents that $\varDelta T_0$ elapses in the clock's rest frame.  
If the observer at rest with 4-velocity $u^{\mu}$ measures this in the observer's rest frame, the elapsed time 
$\varDelta T$ is 
\begin{eqnarray}
    \varDelta T &\equiv& - t_{\mu} u^{\mu} 
    = - \varDelta T_0 \,\bar{u}_{\mu}u^{\mu} = \frac{\varDelta T_0}{\sqrt{1-V^2}}\,,
\end{eqnarray}
where we have used Eq.~\eqref{eq:Lofac}.  Hence, 
\begin{equation}
    \varDelta T_0 = \varDelta T \sqrt{1-V^2} < \varDelta T\,.
\end{equation}
The elapsed time of the moving clock $\varDelta T_0$ is shorter than that of the clock at rest $\varDelta T$.  Clocks in motion tick slowly. 
This is the kinetic time dilation.  It may be often called the special relativistic time dilation.  However, it is also valid in general relativity, because it is derived in covariant way without using the Lorentz transformation.

\subsection{Lorentz contraction}

Suppose that a perfectly rigid rod is placed parallel to the direction $e^{\mu}$ and moving with 4-velocity $\bar{u}^{\mu}$.  
Observer $A$ at rest measures the length of the moving rod and obtain the value of $L$ in $A$'s rest frame.  
The 4-vector 
\begin{equation}
    \ell^{\mu} \equiv L \,e^{\mu}
\end{equation}
represents this fact.  
On the other hand, the ``proper'' length $L_0$ measured in the rod's rest frame is
\begin{equation}
    L_0 \equiv \ell_{\mu} \bar{e}^{\mu} = L \,e_{\mu}\bar{e}^{\mu} = \frac{L}{\sqrt{1-V^2}}\,, 
\end{equation}
where we have used Eq.~\eqref{eq:Lofac}. 
Hence, 
\begin{equation}
    L = L_0 \sqrt{1-V^2} < L_0\,. 
\end{equation}
The length $L$ of the moving rod is shorter than the ``proper'' length $L_0$ measured in the rod's rest frame.  
Rods in motion contract.  
This is known as the Lorentz contraction.  

\subsection{Composition law of 3-dimensional velocities}

We already assumed that observer $A$ with 4-velocity $u^{\mu}$ is ``at rest''  and observer $B$ with 4-velocity $\bar{u}^{\mu}$ is ``moving'' with relative velocity $V$. 
Furthermore, let us assume that observer $C$ is moving with 4-velocity $\tilde{u}^{\mu}$ from $B$ with relative velocity $W$. 
Then, 
\begin{eqnarray}
    \tilde{u}^{\mu} &=& 
    \frac{\bar{u}^{\mu} + W \bar{e}^{\mu}}{\sqrt{1-W^2}}\,. \label{eq:compoutil}
\end{eqnarray}
Using Eqs.~\eqref{eq:B1}, \eqref{eq:B2}, \eqref{eq:compoutil}, and 
eliminating $\bar{u}^{\mu}$ and $\bar{e}^{\mu}$, we obtain 
\begin{equation}
    \tilde{u}^{\mu} = \frac{u^{\mu} + U e^{\mu}}{\sqrt{1 - U^2}} \,,
\end{equation}
where 
\begin{equation}\label{eq:U}
    U \equiv \frac{V + W}{1 + VW}\,.
\end{equation}
This means that observer $C$ is moving from $A$ with relative velocity $U$ of Eq.~\eqref{eq:U}.   
This is well known as the composition law of 3-dimensional velocities.

\subsection{Doppler effect and aberration of light}

The formulae are already derived in Sec. \ref{sec:Dop}.  Here we list the related equations 
Eqs.~\eqref{eq:DopB}, \eqref{eq:DopA}, and 
\eqref{eq:aberration} again in the following: 
\begin{eqnarray}
    \bar{\omega} = \omega \frac{\sqrt{1-V^2}}{1 - V \cos\bar{\vartheta}}\,, \\
    \omega = \bar{\omega} \frac{\sqrt{1-V^2}}{1 + V \cos{\vartheta}}\,, \\
    \cos\bar{\vartheta} = \frac{\cos\vartheta + V}{1 + V \cos\vartheta}\,,  
\end{eqnarray}
where $\omega \equiv - k_{\mu} u^{\mu}$ is the frequency measured by observer $A$ at rest, 
$\bar{\omega} \equiv - k_{\mu} \bar{u}^{\mu}$ is the frequency measured by moving observer $B$, 
${\vartheta}$ is the angle of incidence in observer $A$'s rest frame, 
and $\bar{\vartheta}$ represents the angle of incidence in observer $B$'s rest frame. 
We do not need the Lorentz transformation to derive these formulae.  

\subsection{Transformation of the electromagnetic fields}

Let us define the electromagnetic tensor
\begin{equation}
    F_{\mu\nu} \equiv \partial_{\mu} A_{\nu} - \partial_{\nu} A_{\mu}\,,
\end{equation}
and it's dual tensor 
\begin{equation}
    {}^{\ast}\!F_{\mu\nu} \equiv \frac{1}{2} \varepsilon_{\mu \nu \alpha \beta}\,F^{\alpha \beta}\,,
\end{equation}
where $\varepsilon_{\mu \nu \alpha \beta}$ is the Levi-Civita complete antisymmetric tensor. 

Once the electromagnetic tensor is given, the electric and the magnetic (magnetic flux density) field vectors for observer $A$ are
\begin{eqnarray}
    E_{\mu} &\equiv& F_{\mu\nu} u^{\nu}\,, \\
    B_{\nu} &\equiv& {}^{\ast}\!F_{\mu\nu} u^{\mu}\,.
\end{eqnarray}
Both are spacelike in the sense $E_{\mu} u^{\mu} = B_{\nu} u^{\nu}= 0$.  

We introduce the unit spacelike vector $n^{\mu}$ which satisfies 
$u_{\mu}n^{\mu} = 0, e_{\mu}n^{\mu} = 0, n_{\mu}n^{\mu} = 1$.  
Note that $n^{\mu}$ also satisfies $\bar{u}_{\mu}n^{\mu} = 0, \bar{e}_{\mu}n^{\mu} = 0$, which can be easily proven using Eqs.~\eqref{eq:B1} and \eqref{eq:B2}. Therefore, the spacelike vector $n^{\mu}$ is ``perpendicular'' to both $e^{\mu}$ and $\bar{e}^{\mu}$. 

Using $e^{\mu}$ and $n^{\mu}$, we define the ``parallel'' and the ``perpendicular'' components of the electromagnetic field vectors as follows:
\begin{eqnarray}
    E_{/\!/} &\equiv& e^{\mu} E_{\mu} = e^{\mu} F_{\mu\nu} u^{\nu}\,, \\
    E_{\perp} &\equiv& n^{\mu} E_{\mu} = n^{\mu} F_{\mu\nu} u^{\nu} \,, \\
    B_{/\!/} &\equiv& e^{\nu} B_{\nu} = e^{\nu}\, {}^{\ast}\!F_{\mu\nu} u^{\mu}\,, \\
    B_{\perp} &\equiv& n^{\nu} B_{\nu} = n^{\nu} \,{}^{\ast}\!F_{\mu\nu} u^{\mu}\,.
\end{eqnarray}

As for moving observer $B$ with $\bar{u}^{\mu}$, the ``parallel'' components are
\begin{eqnarray}
    \bar{{E}}_{/\!/} &\equiv& \bar{e}^{\mu} F_{\mu\nu} \bar{u}^{\nu}\\   
    &=& \frac{e^{\mu} + V u^{\mu}}{\sqrt{1-V^2}} F_{\mu\nu} \frac{u^{\nu} + V e^{\nu}}{\sqrt{1-V^2}}\\   
    &=& e^{\mu} F_{\mu\nu}u^{\nu} 
    \\ &=& E_{/\!/}  \,,
\end{eqnarray}
\begin{eqnarray}
    \bar{B}_{/\!/}&\equiv& \bar{e}^{\nu} \,{}^{\ast}\!F_{\mu\nu} \bar{u}^{\mu}\\  
    &=& \frac{e^{\nu} + V u^{\nu}}{\sqrt{1-V^2}}\, {}^{\ast}\!F_{\mu\nu} \frac{u^{\mu} + V e^{\mu}}{\sqrt{1-V^2}}\\   
      &=& e^{\nu}\, {}^{\ast}\!F_{\mu\nu}u^{\mu} \\ &=& B_{/\!/}\,.
\end{eqnarray}
In the same way, the ``perpendicular'' components for moving observer $B$ are
\begin{eqnarray}
    \bar{E}_{\perp} &\equiv&  {n}^{\mu} F_{\mu\nu} \bar{u}^{\nu}\\   
    &=& {n}^{\mu} F_{\mu\nu} \frac{u^{\nu} + V e^{\nu}}{\sqrt{1 - V^2}}\\ 
    &=& \frac{{n}^{\mu} F_{\mu\nu} u^{\nu} + {n}^{\mu} F_{\mu\nu} V e^{\nu}}{\sqrt{1 - V^2}}\\  
    &=& \frac{E_{\perp}+ \boldsymbol{n}\cdot(\boldsymbol{V}\times\boldsymbol{B})}{\sqrt{1 - V^2}} \,, 
\end{eqnarray}
\begin{eqnarray}
    \bar{B}_{\perp} &\equiv& n^{\nu} \,{}^{\ast}\!F_{\mu\nu} \bar{u}^{\mu}\\    
    &=& {n}^{\nu} \,{}^{\ast}\!F_{\mu\nu} \frac{u^{\mu} + V e^{\mu}}{\sqrt{1 - V^2}}\\   
    &=& \frac{{n}^{\nu} \,{}^{\ast}\!F_{\mu\nu} u^{\mu} + {n}^{\nu} \,{}^{\ast}\!F_{\mu\nu} V e^{\mu}}{\sqrt{1 - V^2}}\\   
    &=& \frac{B_{\perp}- \boldsymbol{n}\cdot(\boldsymbol{V}\times\boldsymbol{E})}{\sqrt{1 - V^2}}  \,,
\end{eqnarray}
where 3-dimensional vectors are
\begin{eqnarray}
    \boldsymbol{n} &=& \left(n^1, n^2, n^3\right)\,, \\
    \boldsymbol{V} &=& \left(V e^1, V e^2, V e^3\right)\,, \\
    \boldsymbol{E} &=& \left(F_{10}, F_{20}, F_{30}\right)\,, \\
    \boldsymbol{B} &=& \left(F_{23}, F_{31}, F_{12}\right)
\end{eqnarray}
in the coordinate system $u^{\mu} = (1, 0, 0, 0)$. 

\section{A conventional derivation of Eq.~\eqref{eq:tdcirc} based on the Newtonian analogy}\label{sec:B3}

A conventional derivation\cite{taylorwheeler} of the time dilation of the moving clock on the  circular orbit uses the proper time 
\begin{equation}
    d\tau = \sqrt{-g_{\mu\nu} dx^{\mu} dx^{\nu}} \,.
\end{equation}
The elapsed proper time $d\bar{\tau}$ of the moving clock on the circular orbit of radius $r$ is
\begin{eqnarray}
    d\bar{\tau} &=& 
    \sqrt{\left(1-\frac{r_g}{r} \right)dt^2 - r^2 d\phi^2} \\
    &=& dt \sqrt{\left(1-\frac{2GM}{r}\right) - v^2}  \,,
\end{eqnarray}
where $\displaystyle v \equiv r \frac{d\phi}{dt}$ is the Newtonian velocity of circular motion.  
Based on the Newtonian analogy, we can assume 
the balance of the Newtonian gravity and 
the centripetal force
\begin{equation}
    \frac{GMm}{r^2} = \frac{m v^2}{r}\,,\ \ \therefore\ \ 
    v^2 = \frac{GM}{r}\,. 
\end{equation}
Note that this balance equation  holds only in the case of the Newtonian circular orbit, and it is not applicable to the case of non-circular orbits.
Then, 
\begin{equation}
    d\bar{\tau} = dt \sqrt{\left(1-\frac{2GM}{r}\right) - \frac{GM}{r}} 
    = dt \sqrt{1 - \frac{3}{2}\frac{r_g}{r}} \,.
\end{equation}
On the other hand, the elapsed proper time $d\tau_1$ of the clock at rest at $r=r_1$ is
\begin{equation}
    d\tau_1 = dt \sqrt{1-\frac{r_g}{r_1}}  \,. 
\end{equation}
The ratio
\begin{equation}
    \frac{d\bar{\tau}}{d\tau_1} = \frac{\sqrt{1 - \frac{3}{2}\frac{r_g}{r}}}{\sqrt{1-\frac{r_g}{r_1}}}
\end{equation}
happens to be the same form as Eq.~\eqref{eq:tdcirc}. However, this kind of the Newtonian analogy might necessarily contain some Newtonian approximation, and it is not guaranteed that the resultant equation holds exactly in the Schwarzschild spacetime. 

\section*{ACKNOWLEDGMENTS}

I would like to thank Jorma Louko for the valuable information.





\begin{thebibliography}{9}
\bibitem{ellis69} See Eqs.~$(6.5a), \ (6.9)$ of Ellis\cite{ellis}.  It is originally written as
    \begin{itemize}
        \item[]$k^a\ k_a = 0\ ,$
        \item[]$k_{a;b}\, k^b = 0\ .$
    \end{itemize}
\bibitem{elliseq1}
    See Eq.~$(6.11a)$ of Ellis\cite{ellis}. It is originally written as 
    \begin{itemize}
        \item[]
        $u_a = \cosh \beta\, u_a + \sinh \beta\, e_a, \quad e_a e^a = 1, \quad e^a u_a = 0, $
    ..., and $V = \operatorname{tgh} \beta$ is the velocity of relative motion. (The same $u_a$ appears in the both sides.  It might be a typo. )
    \end{itemize}
\bibitem{ellis612}
    See Eq.~$(6.12)$ of Ellis\cite{ellis}. It is originally written as
    \begin{itemize}
        \item[]
        $k^a = (-u_b\,k^b) (u^a + n^a), \quad n^a\,b_a = 1, \quad n^a\,u_a = 0 \ .$
    \end{itemize}
\bibitem{ellis610b}
    See Eq.~$(6.10b)$ of Ellis\cite{ellis}. It is originally written as
    \begin{itemize}
        \item[]
        $\displaystyle 1 + z  = \frac{(u^{a}k_{a})_{\rm emitter}}{(u^{a}k_{a})_{\rm observer}}$
    \end{itemize}
\end{thebibliography}



\end{document}